# Resummed thermodynamic perturbation theory for bond cooperativity in associating fluids

Bennett D. Marshall and Walter G. Chapman

*Department of Chemical and Biomolecular Engineering, Rice University, 6100 S. Main, Houston, Tx 77005*

We develop a resummed thermodynamic perturbation theory for bond cooperativity in associating fluids by extension of Wertheim's multi-density formalism. We specifically consider the case of an associating hard sphere with two association sites and both pairwise and triplet contributions to the energy. To test the theory we perform new monte carlo simulations. Theory and simulation are found to be in excellent agreement.

Hydrogen bond cooperativity is a common effect observed through experiment[1] and quantum mechanics[2] (QM) calculations. QM has shown that hydrogen fluoride (HF) shows a strong degree of hydrogen bond cooperativity. The binding energy per hydrogen bond $-\varepsilon_{HB}$ increases as cluster size increases until a chain length of approximately six at which point $\varepsilon_{HB}$ stabilizes as chain length increases.[2] Since the system is no longer pairwise additive many of the common theories for associating fluids are no longer applicable. For instance, Wertheim's mult-density formalism[3-5] for associating fluids is built on the assumption of pairwise additivity. To fill this gap a number of lattice theories[6,7] have been developed, as well as the approach of Sear and Jackson (SJ)[8]. The approach of SJ considered bond cooperativity in polymerizing two site associating fluids such as HF. The specific model considered the total energy of a fluid composed of $N_P$ hard spheres of diameter $d$ with two association sites as[8]

$$U(1...N_P) = \frac{1}{2}\sum_{i,j}\left(\phi_{HS}(ij) + \phi_{as}^{(2)}(ij)\right) + \frac{1}{6}\sum_{i,j,k}\phi_{as}^{(3)}(ijk) \qquad (1)$$

where (1) accounts for pairwise $\phi_{as}^{(2)}$ and triplet $\phi_{as}^{(3)}$ association interactions. The pairwise term was treated as conical square wells[9-11] where if two sites overlap the energy of the system is changed by a factor of $-\varepsilon^{(1)}$. In this approach the size (solid angle) of the association site is governed by the angle $\theta_c$ and the range the association by $r_c$. The triplet $\phi_{as}^{(3)}$ term modifies the pair potential by stating that for each molecule bonded twice the energy is changed by an additional $-\left(\varepsilon^{(2)} - \varepsilon^{(1)}\right)$. Using this model SJ[10] developed a theory in the associating ideal gas limit and then used this form of this theory to construct, in an adhoc fashion, the equation of state at higher densities.

For systems which exhibit pairwise additivity, Wertheim's multi – density[3-5] approach has proven to be highly accurate in modeling associating fluids[12-16]. Since Wertheim's theory is founded on the assumption of pairwise additivity, it seems certain the approach would not be applicable to systems which do not obey pairwise additivity. In this letter we show that the multi – density formalism of Wertheim can be applied to systems which are non pair – wise additive. We accomplish this through a new type of perturbation theory, but first, let's review Wertheim's $N^{th}$ order perturbation theory TPT$N$ for two site associating fluids.[5] In TPT$N$ the Helmholtz free energy for the two site case is given by[5]

$$\frac{A - A_{HS}}{k_B TV} = \rho \ln \frac{\rho_o}{\rho} - \sigma_A - \sigma_B + \frac{\sigma_A \sigma_B}{\rho_o} + \rho - \frac{\Delta c^{(o)}}{V} \qquad (2)$$

Here $\rho$ is the total density, $\rho_o$ is the monomer density, $\sigma_A = \rho_A + \rho_o$ where $\rho_A$ is the density of molecules bonded at site $A$. The associative fundamental graph sum $\Delta c^{(o)}$ is given by $\Delta c^{(o)} = \sum_{n=1}^{N} \Delta c_n$ where $\Delta c_n$ is the $n^{th}$ order contribution (involves $n$ association bonds) and is given by

$$\frac{\Delta c_n}{V} = \sigma_A \sigma_B \rho_o^{n-1} I_n \tag{3}$$

The integrals $I_n$ are given by

$$I_n = \frac{1}{\Omega^n} \int f_{AB}(12)...f_{AB}(n,n+1) G(1...n+1) d(2)...d(n+1) \tag{4}$$

The $f_{AB}(12)$ are the association Mayer functions and Wertheim defines the functions $G(1...n+1)$ as, "the subset of graphs in $g(1...n+1)$ such that combining them with the chain produces an irreducible graph; $g(1...s)$ denotes the s particle correlation function of the *reference* system"[5]. This means, for instance, that in a second order perturbation theory the contribution $\Delta c_2$ will include the triplet correlation function $g(123)$, but one must subtract of the contribution from the first order term $\Delta c_1$ to keep from double counting. We then obtain the $G(1...s)$ by summing $g(1...s)$ and all products of $g$'s obtained by partitioning 1...s into subsequences which share the switching point and associating a -1 with each switching point.[5] A few examples include

$$G(12) = g(12)$$
$$G(123) = g(123) - g(12)g(23) \tag{5}$$
$$G(1234) = g(1234) - g(123)g(34) - g(12)g(234) + g(12)g(23)g(34)$$

The general idea of TPT$N$ is then to build up chains by adding in higher order contributions while subtracting off lower order contributions. What does this have to do with bond cooperativity?

Lets consider the case of a 2 site sphere with the sites having a 180 degree bond angle such that there is no steric hindrance between sites. For this case, TPT1, should be sufficient to describe the association.[17] However, now let's consider the case that there is bond cooperativity as given in (1). For this case the first bond in the chain contributes a -$\varepsilon^{(1)}$ to the energy while each subsequent bond in the chain has an energy -$\varepsilon^{(2)}$. Instead of using TPT$N$ to enforce steric constraints we will employ it to correct for bond cooperativity. Following the same logic as discussed above for TPT$N$ we propose the following form for the fundamental graph sum: $\Delta c^{(o)} = \sum_{n=1}^{N} \Delta \tilde{c}_n$ where $\Delta \tilde{c}_n$ is given by (3) with the substitution $I_n \to \tilde{I}_n$ where the integrals $\tilde{I}_n$ are given by

$$\tilde{I}_n = \frac{1}{\Omega^n} \int g(12)...g(n,n+1)\tilde{F}(1...n+1)d(2)...d(n+1) \qquad (6)$$

Where now the $g(12)$ play the part of the $f_{AB}(12)$ in (4) and the $\tilde{F}$'s play the part of the $G$'s. The $\tilde{F}(1...s)$ are obtained by the same partitioning procedure given for the $G(1...s)$ with the $g(1...s)$ exchanged for the functions $\tilde{f}(1...s)$. The $\tilde{f}(1...s)$ give the product of all association Mayer functions in a chain of length $s$. For the specific model considered here, conical square well sites[9-11], this function is given by $\tilde{f}(1...s) = C_s \prod_{k=1}^{s-1} \lambda(k,k+1)$ where $C_s = f_{AB}^{(1)}\left(f_{AB}^{(2)}\right)^{s-2}$ and $f_{AB}^{(j)} = \exp(\varepsilon^{(j)}/k_B T) - 1$. The term $\lambda(12) = 1$ if molecules 1 and 2 are positioned and oriented such that association occurs between site $A$ on 1 and site $B$ on 2 and $\lambda(12) = 0$ otherwise. Now we simplify (6) as $\tilde{I}_n = \Delta^n \Xi_{n+1}$ where $\Delta = \pi(1-\cos\theta_c)^2 \int_d^{r_c} g(r) r^2 dr$ and $\Xi_s$ is obtained by factoring out the $C_s$ in the partitioning of $F(1...s)$ which we obtain as

$$\Xi_m = f_{AB}^{(1)}\left(f_{AB}^{(2)} - f_{AB}^{(1)}\right)^{m-2} \tag{7}$$

Finally, we write the new fundamental graph sum for an $N^{th}$ order perturbation theory for bond cooperativity and let the order of perturbation become infinitely large $N \to \infty$ to obtain

$$\frac{\Delta c^{(o)}}{V} = \sigma_A \sigma_B \sum_{n=1}^{\infty} \rho_o^{n-1} \Delta^n \Xi_{n+1} = \frac{\sigma_A \sigma_B f_{AB}^{(1)} \Delta}{1 - \left(f_{AB}^{(2)} - f_{AB}^{(1)}\right)\rho_o \Delta} \tag{8}$$

Equation (8) is the central result of this paper and is remarkably simple. We have developed a new type of perturbation theory where we use perturbations to correct for bond cooperativity. We then allowed the order of perturbation to become infinitely large allowing for a resummation of all pertubative terms. For the case $\varepsilon^{(2)} = \varepsilon^{(1)}$ the standard first order perturbation theory[5] is recovered. Using (8) we can minimize the free energy (2) with respect to $\sigma_B$ and $\rho_o$ to obtain the mass action equations

$$\frac{\sigma_A}{\rho_o} - 1 = \frac{\sigma_A f_{AB}^{(1)} \Delta}{1 - \left(f_{AB}^{(2)} - f_{AB}^{(1)}\right)\rho_o \Delta} \tag{9}$$

and

$$\frac{\rho}{\rho_o} = \left(\frac{\sigma_A}{\rho_o}\right)^2 + \left(f_{AB}^{(2)} - f_{AB}^{(1)}\right)f_{AB}^{(1)}\left(\frac{\sigma_A \Delta}{1 - \left(f_{AB}^{(2)} - f_{AB}^{(1)}\right)\rho_o \Delta}\right)^2 \tag{10}$$

where $\sigma_A = \sigma_B$ due to symmetry. Using (9) we can simplify the free energy in (2) as

$$\frac{A - A_{HS}}{k_B TV} = \rho \ln \frac{\rho_o}{\rho} - \sigma_A + \rho \tag{11}$$

Combining (9) and (10) we obtain a closed equation for $\rho_o$

$$\rho - \rho_o \left(2 f_{AB}^{(2)} \rho \Delta + 1\right) + \rho_o^2 \left(\rho \left(f_{AB}^{(2)} \Delta\right)^2 + 2\Delta\left(e^{\varepsilon^{(2)}} - e^{\varepsilon^{(1)}}\right)\right) - \rho_o^3 f_{AB}^{(2)} \Delta^2 \left(e^{\varepsilon^{(2)}} - e^{\varepsilon^{(1)}}\right) = 0 \tag{12}$$

Equation (12) is similar to the mass action equation obtained by SJ (Eq. (32) of ref[8]), with the only difference being in the theory due to SJ the Mayor function $f_{AB}^{(2)}$ is replaced by an exponential $f_{AB}^{(2)} \to e^{\varepsilon^{(2)}}$. The advantage of (12) is that the exact high temperature limit $\rho = \rho_o$ is obtained, while in the approach of SJ this limit is obtained approximately. Another interesting limit for the monomer density is for $\varepsilon^{(1)} \to 0$ and $\varepsilon^{(2)} \to \infty$ which gives $\rho_o \to e^{-\varepsilon^{(2)}} / \Delta$. This limit shows that even if the energy of the first bond is zero, long chains can still form for large $\varepsilon^{(2)}$.

To test the theory we perform new monte carlo simulations in the canonical ensemble for molecules which interact with the potential given by (1) with the association sites treated as conical square wells[9-11] with potential parameters $\theta_c = 27°$ and $r_c = 1.1d$. The simulations are performed using standard methodology[18]. The simulations were allowed to equilibrate for $N_p \times 10^6$ configurations and averages were taken over another $N_p \times 10^6$ configurations. A trial configuration was generated by displacing and rotating a molecule. For each simulation we used $N_p = 864$ associating hard spheres. While in general having triplet contributions to the system energy can significantly increase computation time, for the current potential we simply needed to keep track of the number of spheres bonded twice which added little computation time as compared to the pairwise additive system.

In Fig. 1 we compare theoretical and simulation predications for the fraction of molecules bonded $k$ times $X_k$ and the excess internal energy $E^* = E/N_p k_B T$. We consider two general cases. In case I we set $\varepsilon^{(1)} = 7k_B T$ and vary $\varepsilon^{(2)}$ and for case II we fix $\varepsilon^{(2)} = 7k_B T$ and vary $\varepsilon^{(1)}$. For each case we use a density of $\rho^* = \rho d^3 = 0.6$. We begin our discussion with case I. For $\varepsilon^{(2)} = 0$, there is no energetic benefit for a sphere to bond twice which results in $X_2 \to 0$. Increasing $\varepsilon^{(2)}$ we see a steady increase in $X_2$ and the fractions $X_1$ and $X_o$ remain nearly constant until $\varepsilon^{(2)} \sim 5k_B T$ at which point they decline sharply. The excess internal energy also remains approximately constant until $\varepsilon^{(2)} \sim 5k_B T$ and then begins to decrease. Theory and simulation are in near perfect agreement.

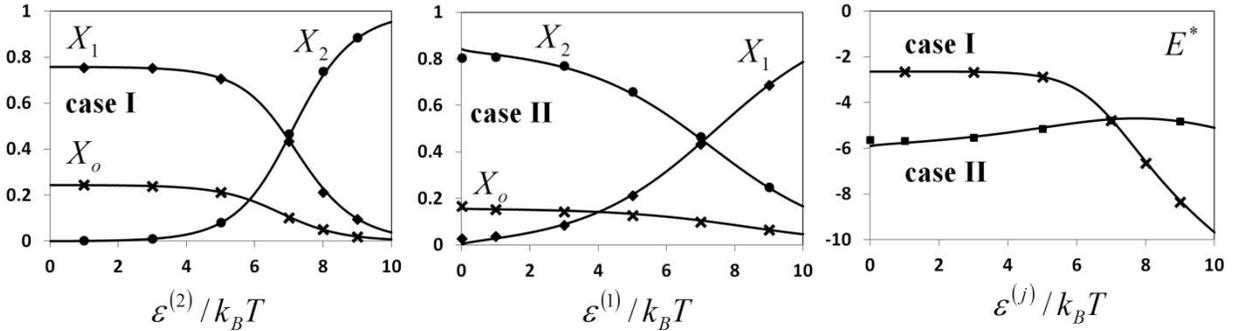

**Figure 1:** Comparison of theoretical predictions (curves) and simulation results (symbols) for the fraction of molecules bonded $k$ times $X_k$ for case I – left panel and case II – center panel. The excess internal energy for both cases is given in the right panel. Each case is at a density of $\rho^* = 0.6$. In case I $\varepsilon^{(1)} = 7k_B T$ and $\varepsilon^{(2)}$ is varied while in case II $\varepsilon^{(2)} = 7k_B T$ and $\varepsilon^{(1)}$ is varied. In the right panel $j = 2$ for case I and $j = 1$ for case II

Now considering case II, set the energy $\varepsilon^{(2)}$ and vary $\varepsilon^{(1)}$. We note the opposite behavior for the fraction $X_2$ as compared to case I. Increasing $\varepsilon^{(1)}$ decreases $X_2$ while increasing $X_1$.

This behavior results from the fact that for small $\varepsilon^{(1)}$ the energy per bond is much lower for the first bond in an associated cluster than all remaining bonds in that cluster. For this reason the system minimizes $X_1$. The behavior of $E^*$ is remarkable for this case. The internal energy increases with increasing $\varepsilon^{(1)}$ until $\varepsilon^{(1)} \sim 7.8 k_B T$ at which point there is a maximum and $E^*$ begins to decrease. It is counter intuitive that increasing $\varepsilon^{(1)}$ could result in an increase in energy. Again, theory and simulation are in excellent agreement. .

We also performed calculations for the monomer fraction $X_o$ and excess internal energy $E^*$ using the theory of SJ[8]; these predictions coincided nearly exactly to the calculations performed with the approach presented in this work. What is unique about the approach developed in this work is that we have shown that bond cooperativity can be treated in the multi – density formalism of Wertheim[3]. This realization has wide ranging impact. It seems that (8) can be paired with contributions for ring formation[15, 17, 19] to obtain a theory for the effect of cooperativity on the balance between chain and ring formation. Also, bond cooperativity can be included in theories which account for multiply bonded sites[20, 21] or possible theories for associating fluids which interact with spherically symmetric potentials[22]. Lastly, Wertheim's theory has found wide application in the field of inhomogeneous associating fluids in the form of density functional theory (DFT);[23-26] the approach presented in this work could be extended in the form of DFT to account for bond cooperativity in inhomogeneous associating fluids.

**Acknowledgments**

The financial support of The Robert A. Welch Foundation Grant No. C – 1241 is gratefully acknowledged